\documentclass[letterpaper, 10 pt, conference]{ieeeconf}  %

\IEEEoverridecommandlockouts                              %

\let\labelindent\relax

\usepackage{amssymb,amsfonts, mathtools}
\usepackage[table]{xcolor}
\usepackage{bm}
\usepackage{enumitem}
\usepackage{siunitx}
\usepackage{makecell}
\usepackage[colorlinks,
  linkcolor=red,
  citecolor=blue, urlcolor=.]{hyperref}
\usepackage[capitalise]{cleveref}
\crefname{assumption}{Assumption}{Assumptions}
\crefname{enumi}{Assumption}{}
\newcommand{\crefdefpart}[2]{%
  \namecref{#1}~\hyperref[#2]{\labelcref*{#1}\ \ref*{#2}}%
}

\usepackage{amsthm}

\newtheoremstyle{ieeeconf}
  {0pt}   %
  {0pt}   %
  {\normalfont}  %
  {\parindent}       %
  {\itshape} %
  {:}         %
  { } %
  {\thmname{#1} \thmnumber{#2}\thmnote{ (#3)}} %
\makeatletter
\renewenvironment{proof}[1][\proofname]{\par
  \pushQED{\qed}%
  \normalfont \topsep\z@
  \trivlist
  \item[\hskip2em
        \itshape
    #1\@addpunct{:}]\ignorespaces
}{%
  \popQED\endtrivlist\@endpefalse
}
\makeatletter

\theoremstyle{ieeeconf}

\newtheorem{assumption}{Assumption}
\newtheorem{definition}{Definition}
\newtheorem{lemma}{Lemma}
\newtheorem{proposition}{Proposition}
\newtheorem{remark}{Remark}
\newtheorem{theorem}{Theorem}

\DeclareMathOperator*{\argmin}{arg\,min}

\newcommand{\x}{\boldsymbol{x}}

\newcommand{\inp}{\boldsymbol{u}}
\newcommand{\w}{\boldsymbol{w}}

\newcommand{\R}{\mathbb{R}}
\newcommand{\N}{\mathbb{N}}
\newcommand{\Prob}{\mathbb{P}}
\newcommand{\E}{\mathbb{E}}
\newcommand{\Q}{\mathbb{Q}}

\newcommand{\B}{\mathbb{B}}
\newcommand{\A}{\mathbf{A}}
\newcommand{\Emat}{\mathbf{E}}
\newcommand{\Bmat}{\mathbf{B}}
\newcommand{\Symm}{\mathbb{S}}

\newcommand{\dd}{\mathop{}\!\mathrm{d}}
\newcommand{\Pset}{\mathcal{P}(\mathcal{Z})}

\newcounter{relctr} %
\everydisplay\expandafter{\the\everydisplay\setcounter{relctr}{0}} %

\newcommand\labelrel[2]{%
  \begingroup
    \refstepcounter{relctr}%
    \stackrel{\textnormal{(\alph{relctr})}}{\mathstrut{#1}}%
    \originallabel{#2}%
  \endgroup
}

\AtBeginDocument{\let\originallabel\label} %

\title{\LARGE \bf Data-driven Distributionally Robust Control Based on Sinkhorn
Ambiguity Sets
}

\author{Riccardo Cescon, Andrea Martin, and Giancarlo Ferrari-Trecate
\thanks{R. Cescon, A. Martin, and G. Ferrari-Trecate
are with the Institute of Mechanical Engineering, EPFL, Switzerland. E-mail addresses: \{riccardo.cescon, andrea.martin, giancarlo.ferraritrecate\}@epfl.ch.}
\thanks{This work was supported as a part of NCCR Automation, a National Centre of Competence in Research, funded by the Swiss National Science Foundation (grant number 51NF40\_225155).}
}

\begin{document}

\maketitle

\thispagestyle{empty}
\pagestyle{empty}

\begin{abstract}
As the complexity of modern control systems increases, it becomes challenging to derive an accurate model of the uncertainty that affects their dynamics. Wasserstein Distributionally Robust Optimization (DRO) provides a powerful framework for decision-making under distributional uncertainty only using noise samples. However, while the resulting policies inherit strong probabilistic guarantees when the number of samples is sufficiently high, their performance may significantly degrade when only a few data are available. Inspired by recent results from the machine learning community, we introduce an entropic regularization to penalize deviations from a given reference distribution and study data-driven DR control over Sinkhorn ambiguity sets. We show that for finite-horizon control problems, the optimal DR linear policy can be computed via convex programming. By analyzing the relation between the ambiguity set defined in terms of Wasserstein and Sinkhorn discrepancies, we reveal that, as the regularization parameter increases, this optimal policy interpolates between the solution of the Wasserstein DR problem and that of the stochastic problem under the reference distribution. We validate our theoretical findings and the effectiveness of our approach when only scarce data are available on a numerical example.

\end{abstract}

\section{Introduction}
Standard stochastic optimal control theory relies on the assumption that a precise statistical description of the noise is available. However, in practice, its probability distribution is often observable only through a finite collection of noise samples. While these can be used to approximate the true distribution, such an estimate can be unsatisfactory for control design as it is heavily based on the quality of the collected data and the available prior knowledge of the unknown variables. When poor statistical information about the uncertainty is used to design controllers, one might face unexpected system behaviors with possible catastrophic out-of-sample performance \cite{kuhn2019wasserstein, kuhn2025distributionally, shafieezadeh2023new}.

Motivated by this observation, Distributionally Robust Optimization (DRO) has emerged as a promising paradigm to deal with uncertainty. It considers a minimax stochastic optimization problem over a neighborhood of the nominal distribution defined in terms of a distance in the space of probabilities. In this way, the solution can be robustified against the most averse distribution which is \textit{close} in some sense to a nominal one, while the degree of conservatism
of the underlying optimization can be modulated by adjusting \textit{how far} the worst distribution can be from the nominal one.
For choosing a proper ambiguity set, one needs to strike a balance between computational tractability and expressivity. In particular, the ambiguity set should be rich enough to cover all relevant distributions, while avoiding pathological distributions that could result in overly conservative decisions.\looseness-1

Several methods have been explored for quantifying the similarity between probability distributions, such as the Kullback–Leibler divergence and the total variation distance \cite{gibbs2002choosing}. Another possibility is to define the ambiguity set with distributions sharing moment information \cite{van2015distributionally}. Among these, recent literature has highlighted the advantages of using metrics related to optimal transport (OT). In particular, ambiguity sets based on the Wasserstein metric \cite{villani2009optimal} provide great expressiveness along with strong statistical out-of-sample guarantees. Thanks to these properties, Wasserstein ambiguity sets have been employed for designing DR control policies. In \cite{taskesen2024distributionally} the authors introduced a generalization of the finite-horizon LQG problem with noise belonging to a Wasserstein ball centered at a Gaussian nominal distribution. Other works focused on the infinite-horizon of the same problem when the nominal distribution is Gaussian \cite{hajar2024distributionally} while \cite{brouillon2025distributionally, kim2023distributional} consider an empirical center. Among other contributions that exploit the Wasserstein metric, \cite{aolaritei2023wasserstein, fochesato2022data, mark2020stochastic, micheli2022data}, %
and \cite{coulson2021distributionally} consider the design of tube-based predictive control schemes. More fundamentally,
\cite{aolaritei2025distributional} provides exact characterizations of how Wasserstein ambiguity sets propagate through the system dynamics.\looseness-1

Despite such encouraging results, the current Wasserstein DRO framework suffers from the following limitations.
First, guaranteeing that the true distribution lies within the Wasserstein ambiguity set with high probability requires choosing a radius that is inversely proportional to the number of available data. If only a few noise samples are available, the corresponding radius may become too large, resulting in overly conservative policies. 
Second, from a modeling perspective, when the true distribution is continuous, Wasserstein DRO might not capture its nature. In fact, \cite{gao2023distributionally} showed that when the nominal distribution has finite support, the worst-case distribution will also be finitely supported.

Motivated by these challenges, this paper makes the following three contributions. First, inspired by recent results in the machine learning and optimization community \cite{azizian2023regularization, blanchet2023unifying, dapogny2023entropy}, we use the Sinkhorn discrepancy \cite{cuturi2013sinkhorn} to define ambiguity sets and accordingly formulate a Sinkhorn DR finite-horizon control problem. Differently from the Wasserstein metric, this allows us to penalize deviations from a reference distribution and thus combine data with prior information about the true distribution. Second, we prove that, despite the additional entropic regularization term in the discrepancy, the Sinkhorn DR linear-quadratic control problem can be expressed as a finite-dimensional convex program when the prior is Gaussian. To do so, we specialize the strong duality result of \cite{sinkhorn} to the case of quadratic loss function and transportation cost given by the Euclidean norm, and leverage the system level parametrization of linear dynamic controllers \cite{wang2019system}.
Last, we derive relations between Sinkhorn and Wasserstein ambiguity sets, showing that the Sinkhorn policy interpolates between the solution of the Wasserstein DR optimal control problem and the $\mathcal{H}_2$ stochastic problem under the reference distribution. These extremes are recovered when the regularization term vanishes or grows unbounded, respectively. These findings are validated with numerical examples.
\newline
\textbf{Notation}. Throughout the paper, given a measurable set $\mathcal{Z}$, we will denote by $\mathcal{P}(\mathcal{Z})$ the set of probability distributions supported on $\mathcal{Z}$. We write $\mu \ll \nu$ to denote that a measure $\mu$ is absolutely continuous with respect to $\nu$. If $\mu, \nu$ are two measures, $\mu\times\nu$ represents the product measure.
Let $[n]$ be the set of indices $\{1,\dots, n\}$. The space of all positive semidefinite matrices of size $d$ is denoted by $\mathbb{S}^d$. We denote by $\|\cdot\|$ the Euclidean norm. Given a positive definite matrix $A\in\R^{d\times d}$, we denote the weighted norm as $\|\cdot\|_A$. The Frobenius norm of a matrix $X\in\R^{m\times d}$ is $\|X\|_F$. The determinant of a square matrix $A$ is denoted by $|A|$. Finally, the notation $\star^{\top}AB$ is short for $B^{\top}AB$.\looseness=-1

\section{Problem Formulation and Preliminaries}
We consider the discrete-time linear time-varying system
\begin{equation}
\label{eq:dyn}
    x_{t+1} = A_t x_{t} + B_t u_{t} + E_t w_{t}\,,
\end{equation}
where $x_{t} \in \mathbb{R}^d, u_{t}\in\mathbb{R}^m, w_{t}\in\mathbb{R}^p$, $A_t\in\mathbb{R}^{d\times d}, B_t\in\mathbb{R}^{d \times m}, E_t\in\mathbb{R}^{d\times p}$ for all $t \in \{0, \dots, N-1\}$, where $N \in \N$ denotes the length of the control horizon.
The evolution of \eqref{eq:dyn}, starting from the initial condition $x_0\in\R^d$, is characterized by the vectors $\x=(x_0^{\top}, \dots, x_{N-1}^{\top})^{\top}$, $\inp=( u_0^{\top}, \dots, u_{N-1}^{\top})^{\top}$, and $\w = (x_0^{\top}, w_0^{\top}, \dots, w_{N-2}^{\top})^{\top}$. With this notation in place, (\ref{eq:dyn}) becomes
\begin{equation}
\label{eq:compact_dynamics}
    \x = Z\A\x + Z\Bmat \inp + \Emat\w\,,
\end{equation}
where $Z$ is the block downshift operator, that is, a block matrix with identity matrices in its first block sub-diagonal and zeros elsewhere, $\A \doteq \text{blkdiag}(A_0, \dots, A_{N-1}, 0_{d\times d})$, $\Bmat \doteq \text{blkdiag}(B_0, \dots, B_{N-1}, 0_{d\times m})$, and $\Emat \doteq \text{blkdiag}(I_{d\times d}, E_0, \dots, E_{N-1})$.

We denote the cost incurred by applying the input sequence $\inp$ in response to the disturbance realization $\w$ by
\begin{align}
\label{eq:cost}
J(\inp, \w) 
&= \begin{bmatrix}
    \x^\top & \inp^\top
\end{bmatrix}
D
\begin{bmatrix}
    \x\\\inp
\end{bmatrix}\,,
\end{align}
where $D\succeq 0$. We also assume that the underlying true noise distribution of $\w$ denoted by $\Prob_*$ is unknown. Instead, we have access to $n\in\N$ i.i.d. samples of noise trajectories of length $N$, $\hat{\w}^i_{[N]} = [(\hat{x}^i_0)^{\top}(\hat{w}^i_0)^{\top} \cdots (\hat{w}^i_{N-2})^{\top}],\ i\in[n]$. %

Consequently, following the DRO literature, we formulate a worst-case cost over a family of probability distributions, and we seek to find the best control action to minimize such a cost. This can be explicitly stated as 
\begin{equation}
\label{eq:objective}
    \min_{\inp}\ \sup_{\Prob\in\mathcal{P}}\ \E_{\w\sim\Prob}[J(\inp, \w)]\,,
\end{equation}
where $J(\inp, \w)$ is the control objective defined in (\ref{eq:cost}), and $\mathcal{P}$ is the ambiguity set. It consists of probability distributions that are related in terms of distance or other suitable properties (e.g. moments) to a reference distribution. In this paper, we use as reference distribution the empirical one induced by the $n$ collected noise trajectories, that is, $\hat{\Prob}_n = 1/n\sum_{i=1}^n \delta_{\hat{\w}^i_{[N]}}$.

For tractability of our formulation and motivated by recent works on DR control \cite{taskesen2024distributionally, lanzetti2024optimality, cescon2025global}, which showed global optimality of linear policies when the nominal distribution is normal, we restrict ourselves to linear time-varying policies. Thus, we consider control actions given by
\begin{equation*}
    u_t = \sum_{k=0}^t K_{t,k}x_k,\ \forall t\in\{0, \dots, N-1\}\,,
\end{equation*}
or, in a more compact form,
\begin{equation}
\label{eq:compact_feedback}
    \inp = \mathbf{K} \x\,,
\end{equation}
with $\mathbf{K}$ being a lower block-triangular feedback matrix.
\begin{remark}[Initial conditions]In the paper, we consider $x_0$ to be uncertain. This is for example the case of the LQG theory where the initial condition is uncertain and only its mean and covariance are known, see, e.g., \cite[Section 5.3]{stengel1994optimal}. However, one can also consider the case where the initial state is given. In this setup the maximization in (\ref{eq:objective}) is carried out only with respect to the disturbance sequence as terms related to a constant $x_0$ can be factored out of the expectation.
\end{remark}

\subsection{System level synthesis (SLS)}
In this section, we provide the necessary background on the SLS approach to optimal controller synthesis by referring the reader to \cite{wang2019system} and \cite{anderson2020system} for a complete discussion. SLS shifts the synthesis problem from the direct design of the controller to the shaping of closed-loop maps from the exogenous disturbance to the state and input signals.

Using (\ref{eq:compact_dynamics}) and (\ref{eq:compact_feedback}), the closed-loop behavior of the system is characterized by the following noise-to-input and noise-to-state maps:%
\begin{align}
\label{eq:SLS-map_x}
    \x &= (I - Z(\A+\Bmat\mathbf{K}))^{-1}\Emat\w = \mathbf{\Phi}_x\w\,,\\
\label{eq:SLS-mapu}
    \inp &= \mathbf{K}(I - Z(\A+\Bmat\mathbf{K}))^{-1}\Emat \w = \mathbf{\Phi}_u\w\,.
\end{align}
The maps $\mathbf{\Phi}_x$ and $\mathbf{\Phi}_u$ are the \textit{system responses} induced by the controller $\mathbf{K}$. We highlight that these operators inherit a block-triangular causal structure. Although (\ref{eq:SLS-map_x}) and \eqref{eq:SLS-mapu} are non-convex in $\mathbf{K}$, they are linear functions of $\{\mathbf{\Phi}_x, \mathbf{\Phi}_u\}$. Therefore, the idea is to optimize directly over these maps: to do so, one must show that there exists a controller $\mathbf{K}$ such that $\x = \mathbf{\Phi}_x\w$ and $\inp = \mathbf{\Phi}_u\w$. From \cite[Theorem 3.1]{anderson2020system}, this is true if and only if 
\begin{equation}
\label{eq:achievability}
    \begin{bmatrix}
        I-Z\A && -Z\Bmat
    \end{bmatrix}
    \begin{bmatrix}
       \mathbf{\Phi}_x\\
       \mathbf{\Phi}_u
    \end{bmatrix} = \Emat\,;
\end{equation}
we call pairs $\{\mathbf{\Phi}_x, \mathbf{\Phi}_u\}$ that satisfy (\ref{eq:achievability}) \textit{achievable}. Moreover, any block-lower-triangular matrices $\{\mathbf{\Phi}_x, \mathbf{\Phi}_u\}$ satisfying (\ref{eq:achievability}) are achieved by the corresponding controller $\mathbf{K} = \mathbf{\Phi}_u\mathbf{\Phi}_x^{-1}$.

\subsection{Characterization of the uncertainty}
In this section, we describe the types of ambiguity sets that are used in the remainder of the paper. We start by recalling some useful definitions from \cite{kuhn2025distributionally, sinkhorn}.
\begin{definition}[Transportation Cost Function]A lower semi-continuous function
$c(x, y): \mathcal{Z} \times \mathcal{Z} \rightarrow \R_+$ that satisfies the identity of indiscernibles (i.e. $c(x,y) = 0$ if and only if $x=y$) is a transportation cost function.
\end{definition}
\begin{definition}[OT discrepancy]The optimal transport discrepancy $OT_c: \Pset \times \Pset \rightarrow [0, +\infty]$ associated with any given transportation cost function $c$ is defined through 
\begin{equation}
\label{eq:OT}
    OT_c(\Prob, \Q) = \inf_{\gamma\in\Gamma(\Prob, \Q)}\E_\gamma[c(x, y)]\,,
\end{equation}
where $\Gamma(\Prob, \Q)$ represents the set of all couplings $\gamma$ between $\Prob$ and $\Q$, that is, all joint probability distributions with marginals $\Prob$ and $\Q$.
\end{definition}

By adding a regularizer to the previous formulation, one obtains the so-called entropy-regularized OT or Sinkhorn discrepancy.
\begin{definition}[Sinkhorn discrepancy]Consider the probability distributions $\Prob, \Q\in\mathcal{P}(\mathcal{Z})$, and let $\mu, \nu$ be reference probability measures over $\mathcal{Z}$ such that $\Prob \ll \mu$ and $\Q \ll \nu$. For a given transport cost $c$ and regularization parameter $\epsilon \geq 0$, the Sinkhorn discrepancy between $\Prob$ and $\Q$ is defined as\looseness=-1
\begin{equation}
\label{eq:sinkhorn}
   W_c^{\epsilon}(\Prob, \Q) = \inf_{\gamma \in \Gamma(\Prob, \Q)}\left\{\E_\gamma [c(x, y)] + \\ 
   \epsilon H(\gamma|\mu\times\nu )\right\}\,,
\end{equation}
where $H(\gamma|\mu\times\nu )$ represents the KL divergence of $\gamma$ with respect to the product measure $\mu\times\nu$:\looseness=-1 
\begin{equation*}
    H(\gamma|\mu\times\nu) = \E_\gamma\left[\log\left(\frac{\dd\gamma(x, y)}{\dd\mu(x)\dd\nu(y)}\right)\right]\,.
\end{equation*}
\end{definition}

We note that any possible choice of $\mu$ in (\ref{eq:sinkhorn}) is equivalent up to a constant. Since in DRO applications the nominal distribution is known and fixed, without loss of generality, we will choose $\mu = \Prob$ in the sequel. Note also that the discrepancy $W_c^{0}$ coincides with the OT discrepancy in (\ref{eq:OT}). Consequently, if the cost function $c$ is defined as the $p$-th power of some metric in the space $\mathcal{Z}$, we retrieve the well-known $p$-th power of the Wasserstein distance between $\Prob$ and $\Q$, e.g., \cite[Definition 6.1]{villani2009optimal}. Intuitively, in (\ref{eq:sinkhorn}), when the regularization parameter $\epsilon$ increases there is an additional constraint on how the mass can be transported from $\Prob$ to $\Q$ based on the reference measure $\nu$. In \cite{sinkhorn}, the authors suggest selecting the reference measure $\nu$ as the Gaussian measure if the underlying true distribution is expected to be continuous, or the counting measure if it is discrete. 

Inherently, the quantities $OT_c(\Prob, \Q)$ and $W_c^{\epsilon}(\Prob, \Q)$ measure how different the probability distributions $\Prob$ and $\Q$ are. They also naturally provide a definition of uncertainty in the space of probability distributions. Specifically, we define the Sinkhorn ambiguity set, denoted as $\mathcal{S}$-set for brevity, of radius $\rho$ and centered at $\Prob$ by  
\begin{align*}
\B_{\rho, \epsilon}(\Prob) &\doteq \{ \Q\in\mathcal{P}(\mathcal{Z}):W_c^{\epsilon}(\Q, \Prob) \leq \rho \} \subset \mathcal{P}(\mathcal{Z})\,.
\end{align*}
Analogously, we can denote the OT ambiguity set by $\B_{\rho}(\Prob)$.
In words, $\mathbb{B}_{\rho, \epsilon}(\Prob)$ contains all probability distributions that are $\rho$ close to $\Prob$ in the Sinkhorn discrepancy.

\section{Sinkhorn Distributionally Robust Control}

In the following section, we address the finite-horizon DR control problem using the Sinkhorn discrepancy presented in the previous section. We first prove key properties of $\mathcal{S}$-sets. Then, we present the main theorem of our work and conclude the section with its proof. 

The next proposition relates $\mathcal{S}$-sets with different radii, as well as with OT-sets. We defer its proof to Appendix \ref{appendix: relationships}.\looseness-1
\begin{proposition}
\label{prop: relationships}
Let $\Prob \in\mathcal{P}(\mathcal{Z})$ and fix the radius $\rho \geq 0$. Then, the following relationships hold:
\begin{enumerate}
\item $\B_{\rho, \epsilon}(\Prob) \subseteq \B_{\rho}(\Prob) \ \forall \epsilon \geq 0\,;$
\item $\B_{\rho, \epsilon_2}(\Prob) \subseteq \B_{\rho, \epsilon_1}(\Prob) \ \forall \epsilon_1, \epsilon_2 : 0\leq \epsilon_1 \leq \epsilon_2\,;$
\item If $\int_{\mathcal{Z}\times\mathcal{Z}} c(x, y)d\Prob(x) d\nu(y) \leq \rho$, then $\B_{\rho, \infty}(\Prob)$ is the singleton $\{\nu\}$. Otherwise, it is the empty set.
\end{enumerate}
\end{proposition}
As hinted by point 3 of the above proposition, the $\mathcal{S}$-set can be empty. As a consequence, the stochastic program in (\ref{eq:objective}) with $\mathcal{S}$-set can become infeasible. Intuitively, for a large enough regularization parameter $\epsilon$ the uncertainty set will only contain distributions that are close to the reference $\nu$, while the first term of the distance will tend to have distributions that are close to the center. Therefore, it might be that, for some combinations of radius and regularizer, the ambiguity set is empty. See Lemma \ref{thm:main} for an exact characterization of these conditions. This is in contrast to the Wasserstein worst-case cost, for which feasibility of \eqref{eq:objective} is always guaranteed. Indeed, when the radius of the Wasserstein ball diminishes, it eventually shrinks to a singleton containing only the center distribution. %

We now shift our attention to the DR control design with the Sinkhorn discrepancy. Using the SLS framework presented in the preliminary section, the control cost we seek to minimize is\looseness-1
\begin{equation*}
    J(\w) = \w^{\top}\mathbf{\Phi}^{\top} D \mathbf{\Phi} \w\,,
\end{equation*}
with $\mathbf{\Phi}$ being the column concatenation of $\mathbf{\Phi}_x$ and $\mathbf{\Phi}_u$.
In the setup where the uncertainty set $\mathcal{P}$ in (\ref{eq:objective}) is $\B_{\rho, \epsilon}(\hat{\Prob}_n)$, and the objective function is $J(\w)$, the problem we want to solve is to find the optimal map $\mathbf{\Phi}$ given by
\begin{equation}
\label{eq:map}
    \mathbf{\Phi}^{\star} \in \argmin_{\mathbf{\Phi}}\ \sup_{\Prob\in\B_{\rho, \epsilon}(\hat{\Prob}_n)}\ \E_{\w\sim\Prob}\Bigl[ \w^{\top}\mathbf{\Phi}^{\top} D \mathbf{\Phi} \w \Bigr]\,,
\end{equation}
while satisfying the achievability constraints in (\ref{eq:achievability}).
The inner maximization in (\ref{eq:map}) is computationally difficult to solve, since it involves an optimization over a possibly infinite number of distributions. The following theorem reformulates this problem into a finite convex program that can be solved with off-the-shelf numerical optimization solvers. Our result explicitly accounts for the entropic regularization term introduced by the Sinkhorn discrepancy in \eqref{eq:sinkhorn} and recovers previously known results for Wasserstein DR control \cite{brouillon2025distributionally} when $\epsilon\rightarrow0$.

\begin{theorem}
\label{thm:main}
Assume $\mathcal{Z} = \R^s$ and $c(x, y) = \|x -y\|^2$. Let $s = d + (N-1)p$ and the reference measure $\nu$ be a multidimensional Gaussian distribution\footnote{Our results can in principle be extended to any reference distribution $\nu$ satisfying \crefdefpart{assumption}{assumption1}-\ref{assumption2}. Exact reformulations of \eqref{eq:map}, however, require case-by-case computations.} with mean vector $m\in\R^s$ and covariance matrix $\Sigma \in\mathbb{S}^s$, that is
\begin{equation*}
\dd\nu(\xi) \! = \! C_s^{-1}\exp\left(\!-\frac{1}{2}(\xi-m)\!^{\top}\Sigma^{-1}(\xi-m)\!\right)\dd\lambda^s(\xi)\,,
\end{equation*}
with $\lambda^s$ the usual $s$-dimensional Lebesgue measure and $C_s = \sqrt{(2\pi)^s|\Sigma|}$.
Then, problem (\ref{eq:map}) is feasible if and only if
\begin{align}
\rho \geq &\frac{\epsilon}{2} \log\left|\Sigma + \frac{\epsilon}{2}I\right| - \frac{\epsilon s}{2}\log\left(\frac{\epsilon}{2}\right) + \frac{\epsilon}{2}\|m\|^2_{\Sigma^{-1}} + \|\hat{\w}^i_{[N]}\|^2
\nonumber\\
-& \frac{1}{n} \sum_{i=1}^n \star^{\top}\left(I +\frac{\epsilon}{2}\Sigma^{-1}\right)^{-1}\left(\hat{\w}^i_{[N]} + \frac{\epsilon}{2}\Sigma^{-1}m\right).\label{eq:feasibility_condition_control}
\end{align}
Moreover, the optimal closed-loop map $\mathbf{\Phi}^{\star}$ in (\ref{eq:map}) is given by the minimizer of
\begin{subequations}
\label{eq:convex SLS}
\begin{align}
&\inf\ \lambda\rho + \frac{1}{n}\sum_{i=1}^n s_i
\nonumber
\\
&\textrm{subject to} ~ \ \forall i\in[n]:\nonumber
\\
&\lambda\in\R_+, Q\in \Symm^s, s_i\in\R, \zeta_i\in\R, \bm{\Phi}\in\R^{N(d+m)\times s}\,, \nonumber
\\
& M = \lambda\left(I+\frac{\epsilon}{2}\Sigma^{-1}\right) - Q, M \succ 0\label{eq:M>0}\,,
\\
&\frac{\lambda\epsilon s}{2}\log\left(\!\frac{\lambda\epsilon}{2}\!\right) - \frac{\lambda\epsilon}{2}\log|\Sigma|- \frac{\lambda\epsilon}{2}\log|M| + \zeta_i\leq s_i\,,\label{eq:logdet inequality}
\\
& \begin{bmatrix}
M && \lambda \hat{\w}^i_{[N]} + \frac{\lambda\epsilon}{2}\Sigma^{-1}m\\
\star && \zeta_i + \lambda\|\hat{\w}^i_{[N]}\|^2 + \frac{\lambda\epsilon}{2}\|m\|^2_{\Sigma^{-1}}\end{bmatrix}\succeq 0\,,\label{eq:LMI2}\\
& \begin{bmatrix}
\label{eq:LMI1 Schur}
Q && \star\\
D^\frac{1}{2}\mathbf{\Phi} && I
\end{bmatrix}\succeq 0\,,\\
& \mathbf{\Phi}\ \textrm{satisfying}\ (\ref{eq:achievability})\nonumber.
\end{align}
\end{subequations}
\end{theorem}
Before presenting the proof, we note that condition \eqref{eq:feasibility_condition_control} can always be satisfied by choosing a sufficiently large radius $\rho$, and we formulate a proposition stating that the optimization problem in (\ref{eq:convex SLS}) is convex. In fact, the objective function is linear in the optimization variables; therefore, \eqref{eq:convex SLS} is a convex program if and only if its feasible set is convex. We report the proof of this proposition in Appendix \ref{app:proof_convexity}.
\begin{proposition}
\label{proposition:convexity}
Consider the optimization problem in (\ref{eq:convex SLS}). The constraints are convex sets in the optimization variables $(\lambda, Q, \mathbf{\Phi}, s_i, \zeta_i),\ \forall i\in[n]$.
\end{proposition}
Note that \eqref{eq:convex SLS}, due to the nonlinear constraint \eqref{eq:logdet inequality}, is a conic problem when $\epsilon\neq0$. This is in contrast with the Wasserstein counterpart of \eqref{eq:convex SLS}, which is a tractable semidefinite program \cite{kuhn2019wasserstein}. This additional complexity is needed to incorporate priors in the synthesis process through an entropic regularization with respect to a reference distribution.
\subsection{Auxiliary results and proof of the main theorem}
Next, we present the technical results that will allow us to give the proof of the main theorem at the end of this section. We start by recalling a strong duality result first proved in \cite{sinkhorn}. Given a loss function $\ell: \R^d\rightarrow\R$, the following worst-case risk
\begin{equation}
\label{eq:worst-case risk}
\sup_{\Prob\in\B_{\rho, \epsilon}(\hat{\Prob}_n)}\ \E_{z\sim\Prob}[\ell(z)]
\end{equation}
admits the strong dual reformulation
\begin{equation}
\label{eq:dual}
    \inf_{\lambda\geq0}\ \Biggl\{\! \lambda\rho + \lambda\epsilon\E_{x\sim\hat{\Prob}_n}\biggl[\log\E_{z\sim\nu}\Bigl[e^{(\ell(z)-\lambda c(x,z))/(\lambda\epsilon)}\Bigr]\biggr]\!\Biggr\}
\end{equation}
under the following assumption:
\begin{assumption}
The reference measure $\nu$, the transport cost $c(x,y)$, the function $\ell$, and the joint distribution $\gamma$ satisfy:
\label{assumption}
\begin{enumerate}[label=(\roman*)]
    \item $\nu\{z:0 \leq c(x, z) < \infty\} = 1$ for $\hat{\Prob}_n$-almost every $x$;\label{assumption1}
    \item $\E_{z\sim\nu}\left[e^{-c(x, z)/\epsilon}\right] < \infty$ for $\hat{\Prob}_n$-almost every $x$;\label{assumption2}
    \item the function $\ell$ is measurable;\label{assumption3}
    \item For every joint distribution $\gamma$ on $\mathcal{Z}\times\mathcal{Z}$ with first marginal distribution $\hat{\Prob}_n$, it has a regular conditional distribution\footnote{We refer to \cite[Chapter 5]{kallenberg2002foundations} for the concept of regular conditional distribution.} $\gamma_x$ given the value of the first marginal equal to $x$.\label{assumption4}
\end{enumerate}
\end{assumption}
\begin{remark}
    
   It can be shown that the worst-case distribution shares the same support as the measure $\nu$. When the true distribution is known to be continuous, choosing $\nu$ as a continuous measure in $\R^d$ can reduce conservatism with respect to Wasserstein DR control. This is because the worst-case distribution in Wasserstein DRO problems is always finitely supported on at most $n+1$ points when the center $\hat{\Prob}$ is the empirical distribution over $n$ points \cite{gao2023distributionally}.
\end{remark}

Motivated by control applications, the following lemma tailors \eqref{eq:dual} to the case of quadratic transport costs and loss functions. The proof can be found in Appendix \ref{appendix: duality}.
\begin{lemma}[Strong dual reformulation for quadratic loss and transport cost]
\label{thm: duality}
Under the same assumptions of Theorem \ref{thm:main}, if $\ell(z) = z^\top Qz + 2q^\top z$ with $Q\in\Symm^d$, $q\in\R^d$ then problem (\ref{eq:worst-case risk}) is feasible if and only if condition \eqref{eq:feasibility_condition_control} holds
and the optimal value of (\ref{eq:worst-case risk}) coincides with the optimal value of the following convex optimization problem
\scalebox{.92}{\parbox{\columnwidth}{%
\begin{align}
\label{eq:convex program}
\inf\ &\lambda\rho + \frac{1}{n}\sum_{i=1}^n s_i
\\
\textrm{s.t.}\ & s_i\in\R, \lambda\in\R_+, \lambda\left(I+\frac{\epsilon}{2}\Sigma^{-1}\right) \succ Q, \quad \forall i\in[n]
\nonumber\\
& \frac{\lambda\epsilon d}{2}\log\left(\frac{\lambda\epsilon}{2}\right) - \frac{\lambda\epsilon}{2}\log|\Sigma|- \frac{\lambda\epsilon}{2}\log\left|\lambda \left(I + \frac{\epsilon}{2}\Sigma^{-1}\right) - Q\right|+
\nonumber\\
&+ \star^{\top}\left(\lambda\left(I +\frac{\epsilon}{2}\Sigma^{-1}\right)-Q\right)^{-1}\left(q+\lambda\left(\hat{\xi}_i+\frac{\epsilon}{2}\Sigma^{-1}m\right)\right)+
\nonumber\\
&- \lambda \|\hat{\xi}_i\|^2 - \frac{\lambda\epsilon}{2}\|m\|^2_{\Sigma^{-1}}\leq s_i\,.\nonumber
\end{align} 
}}
\end{lemma}
\begin{remark}[Comparison with Wasserstein DRO]
A simple comparison with the dual reformulation of the Wasserstein DRO counterpart in \cite[Theorem 11]{kuhn2019wasserstein} shows that our formulation has some additional terms depending on $\epsilon$ which comes from the regularization part of the distance. As we would expect, when $\epsilon\rightarrow 0$ we retrieve the same dual problem of \cite{kuhn2019wasserstein}.\looseness=-1
\end{remark}
Finally, we present the proof of Theorem \ref{thm:main}.

\begin{proof}
We apply Lemma~\ref{thm: duality} with $Q = \mathbf{\Phi}^{\top} D\mathbf{\Phi}$ and $q = 0$. The feasibility condition \eqref{eq:feasibility_condition_control} was already proved in Lemma \ref{thm: duality}. We then focus on deriving \eqref{eq:convex SLS}. We first introduce the auxiliary variables $\zeta_i$, for $i\in[n]$ to upper bound the linear part of the constraint in (\ref{eq:convex program}). This directly yields the constraint \eqref{eq:logdet inequality}. We then consider the remaining linear part of the constraint
\begin{align*}
\lambda^2\star^{\top}\left(\lambda\left(I +\frac{\epsilon}{2}\Sigma^{-1}\right)-Q\right)^{-1}\left(\hat{\w}^i_{[N]}+\frac{\epsilon}{2}\Sigma^{-1}m\right)
\\
- \lambda \|\hat{\w}^i_{[N]}\|^2 - \frac{\lambda\epsilon}{2}\|m\|^2_{\Sigma^{-1}}\leq \zeta_i\label{eq:second_ineq}\,.
\end{align*}
Applying the Schur's complement to the constraint above we obtain \eqref{eq:LMI2}. 
We are left with enforcing $Q = \mathbf{\Phi}^{\top} D\mathbf{\Phi}$. The inequality
\begin{equation*}
    Q \succeq \mathbf{\Phi}^{\top} D\mathbf{\Phi}
\end{equation*} 
can be equivalently reformulated as in \eqref{eq:LMI1 Schur} via Schur's complement. In contrast, the reverse inequality  
\begin{equation*}
    Q \preceq \mathbf{\Phi}^{\top} D\mathbf{\Phi}
\end{equation*}  
does not yield a convex constraint on \(\mathbf{\Phi}\). Importantly, we show that imposing this reverse inequality is not necessary to ensure the optimality of the resulting closed-loop map \(\mathbf{\Phi}\). To show this, let us define the optimization problem in (\ref{eq:convex SLS}) as $P_{ineq}$ and the same problem with the additional constraint $Q \preceq \mathbf{\Phi}^{\top} D\mathbf{\Phi}$ as $P_{eq}$. Since $P_{ineq}$ has a larger feasible set, its solution can be smaller than that of $P_{eq}$. We show that this is not the case.
\\
Assume $\{ Q^{\star}, s^{\star}, \lambda^{\star}, \zeta^{\star}, \mathbf{\Phi}^{\star}\}$ to be optimal for $P_{ineq}$ with the optimal cost given by $J^{\star} = \lambda^{\star}\rho + \frac{1}{n}\sum_{i=1}^n s^{\star}_i$. We can construct a candidate solution for $P_{eq}$ as $\{\mathbf{\Phi}^{\star \top} D\mathbf{\Phi}^{\star}, s^{\star}, \lambda^{\star}, \zeta^{\star}, \mathbf{\Phi}^{\star}\}$. This satisfies the equality constraint by design. Note also that it has the same cost $J^{\star}$ since the latter depends only on $\lambda^{\star}$ and $s^{\star}$. Therefore, if such a candidate point is feasible, it is also optimal. 
\\
To show feasibility, let us further define $\Tilde{Q} = \mathbf{\Phi}^{\star \top} D\mathbf{\Phi}^{\star}$. By construction, $\Tilde{Q} \preceq Q^{\star}$. We proceed to verify that (\ref{eq:logdet inequality}) and (\ref{eq:LMI2}) are satisfied also with $\Tilde{Q}$. Since the logarithm is a monotone function and given that $A \preceq B$ implies $ |A| \leq  |B|$, we deduce that 
\begin{equation*}
\scalebox{.99}{$
\frac{\lambda\epsilon}{2}\log\left|\lambda \left(I + \frac{\epsilon}{2}\Sigma^{-1}\right) - Q^{\star}\right| \leq \frac{\lambda\epsilon}{2}\log\left|\lambda \left(I + \frac{\epsilon}{2}\Sigma^{-1}\right) - \Tilde{Q}\right|.
$}
\end{equation*} 
Hence, (\ref{eq:logdet inequality}) is verified when using $\Tilde{Q}$.
By using a similar reasoning for the term $\lambda \left(I + \frac{\epsilon}{2}\Sigma^{-1}\right) - Q$ one obtains $\lambda \left(I + \frac{\epsilon}{2}\Sigma^{-1}\right) - Q^{\star} \preceq \lambda \left(I + \frac{\epsilon}{2}\Sigma^{-1}\right) - \Tilde{Q}$, which implies that (\ref{eq:LMI2}) is also verified. Hence, $\{ \Tilde{Q}, s^{\star}, \lambda^{\star}, \zeta^{\star}, \mathbf{\Phi}^{\star}\}$ is a feasible solution and the proof is concluded.
\end{proof}

\section{Numerical Results}
We now numerically validate the theoretical results of Proposition~\ref{prop: relationships} and showcase the effectiveness of using the Sinkhorn discrepancy when only a few uncertainty samples are available for control design. In the experiments, we consider a discrete-time stochastic mass–spring-damper system described
by the linear dynamics:
\begin{equation*}
   x_{t+1} = \begin{bmatrix}
        1 && T_s \\
        -\frac{kT_s}{m} && 1-\frac{cT_s}{m}
    \end{bmatrix}x_t + \begin{bmatrix}
        0 \\
        \frac{T_s}{m}
    \end{bmatrix}u_t + w_t\,,
\end{equation*}
with mass $m = \SI{1}\kilogram$, spring and damping constants $k = \SI{1}{\newton\per\meter}$ and $c = \SI{1}{\newton\per\meter\second}$, respectively, and sampling time $T_s = \SI{1}{\second}$.\looseness-1\footnote{All our experiments were run on an M3 Pro CPU machine with 36GB RAM. All SDP problems are modeled in Matlab 2023a using Yalmip and solved with MOSEK. Our source code is publicly available at \href{https://github.com/DecodEPFL/Sinkhorn_DRC.git}{\texttt{https://github.com/DecodEPFL/Sinkhorn\_DRC.git}}.}
\newline
\textit{1) Validity of points 1-3 of Proposition \ref{prop: relationships}}. In our first test, we synthesize robust controllers starting from a batch of sampled noise trajectories, and we verify the relationships of the proposition. Specifically, we select different values of the regularizer $\epsilon$ and corresponding radii $\rho$ ensuring the feasibility of the problem according to \eqref{eq:feasibility_condition_control}. We compare the behavior of the Sinkhorn worst-case cost with the Wasserstein counterpart. The $n = 20$ noise samples are taken from a Gaussian distribution with zero mean and covariance matrix $0.5I$. The considered horizon length is $N=10$. The selected radii ensuring feasibility are $\rho\in\{3.5, 4, 5\}$. The reference measure $\nu$ in Theorem \ref{thm:main} was chosen as a multidimensional normal with zero mean and covariance matrix $0.1I$.

\begin{figure}[th!]
   \centering
   \includegraphics[trim=0 0 0 0, clip, width=0.99\linewidth]{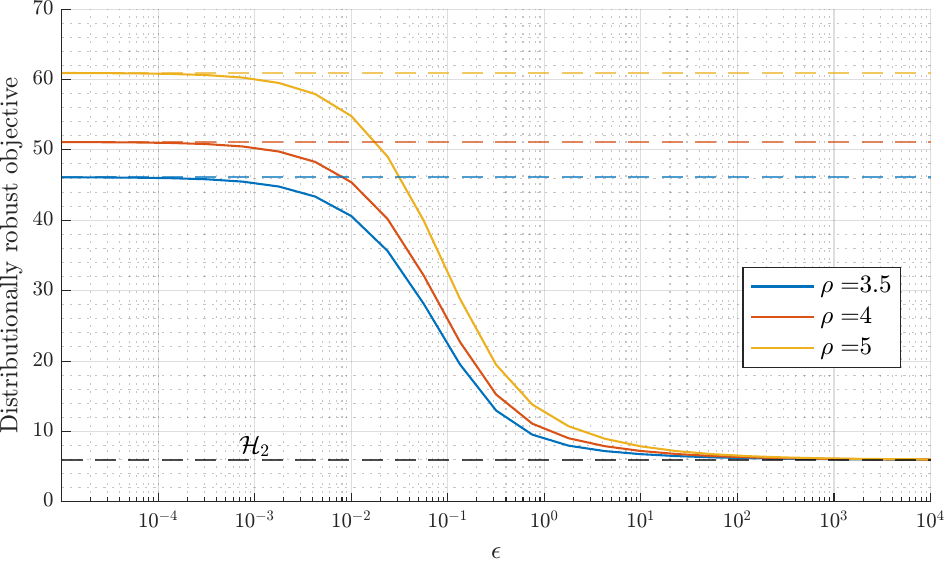}
    \caption{Comparison of worst-case costs between Wasserstein (dashed) and Sinkhorn (solid) for different values of ambiguity set radius $\rho$ and regularization parameter $\epsilon$. The black dashed line shows the optimal expected cost under the prior distribution $\nu$.}
   \label{fig:epsilon_vs_radii}
\end{figure}

In Figure \ref{fig:epsilon_vs_radii} we can see that, as expected, the worst-case costs between Wasserstein and Sinkhorn coincide when the regularization parameter tends to zero. Indeed, we can see that for each value of the radius the solid and dashed lines overlap for small regularizer. Furthermore, we also see that the Sinkhorn worst-case cost is a monotonically decreasing function of $\epsilon$. Again, this is expected from the proposition, since the $\mathcal{S}$-set shrinks with $\epsilon$ and the worst-case cost can only be smaller or equal over a smaller ball. Finally, also the third point of Proposition~\ref{prop: relationships} is verified given that for each possible value of radius $\rho$ the Sinkhorn worst-case cost converges to a value that corresponds to the $\mathcal{H}_2$ controller cost. Indeed, if the problem is feasible, the Sinkhorn ambiguity set converges to a singleton containing only the reference measure $\nu$ which is Gaussian.
\newline
\textit{2) Advantages of Sinkhorn DR control}. In this second test, we show the advantages of using the Sinkhorn discrepancy to describe the uncertainty in the samples. In particular, we envision our approach to be beneficial when the number of samples to build the ambiguity set is limited. Indeed, in such scenarios, the introduction of the regularizer represents a sort of prior on the true distribution. To show this, we consider only $n=4$ noise trajectories extracted from a Gaussian distribution with zero mean and $0.3I$ covariance matrix, horizon $N=15$, and radii $\rho \in\{3, 20\}$. The selected regularization parameters are $
\epsilon\in\{0.001, 0.01, 0.1\}$. We also consider the reference measure $\nu$ used in the previous experiment. In this setup, we compute the optimal maps $\mathbf{\Phi}^{\star}$ by solving DR optimal control problems with Wasserstein and Sinkhorn ambiguity sets for the different radii and regularizers along with the nominal and $\mathcal{H}_2$ maps. We then compare the performance of such control schemes when the noise is generated by the true distribution. To do so, we compute $\E_{\w\sim\Prob_*}\Bigl[ \w^\top\mathbf{\Phi}^{\star \top} D \mathbf{\Phi}^{\star} \w \Bigr]$ which, since the control objective is quadratic, is equivalent to $\|D^{\frac{1}{2}}\mathbf{\Phi}^{\star}\Sigma_{\w}^{\frac{1}{2}}\|^2_F$ with $\Sigma_{\w}$ being the covariance of the noise under $\Prob_*$.
The results obtained are reported in \cref{tab:cost_comparison} for radii $\rho = 3$ and $\rho = 20$. We can see that DR approaches are beneficial since both Sinkhorn and Wasserstein obtain a cost which is smaller than the nominal one of $21.12$, which exploits only the limited number of samples. In addition, Sinkhorn performs better than Wasserstein since it exploits the prior information encapsulated in the reference $\nu$ and not just the noise samples. This is the case even though the prior was chosen not exactly as the underlying true distribution. For completeness, we also computed the cost we would have incurred if the distribution were known, i.e., the $\mathcal{H}_2$ controller cost, which is $9.33$. This value is obviously smaller than those appearing in \cref{tab:cost_comparison} since it assumes perfect knowledge of the disturbance distribution. However, the value is not significantly smaller than those obtained with the robust approaches.
\begin{table}[ht]
\centering
\begin{tabular}{|c|c|c|c|c|c|}
\hline
Controller type & \makecell{Sinkhorn, \\ $\epsilon = 0.01$} & \makecell{Sinkhorn, \\ $\epsilon = 0.05$} & \makecell{Sinkhorn, \\ $\epsilon = 0.1$} & Wasserstein \\ \hline
Cost, $\rho = 3$  & 10.91 & 10.87 & 10.18 & 11.03 \\ \hline
Cost, $\rho = 20$ & 10.77 & 10.76 & 10.67 & 10.80 \\ \hline
\end{tabular}
\caption{Comparison of realized costs for different controllers and values of $\rho$ under the true unknown distribution.}
\label{tab:cost_comparison}
\end{table}

\section{Conclusion}
We have presented a novel method to design DR policies for finite-horizon control using the Sinkhorn discrepancy. To do so, we first leveraged a strong duality result and combined it with the SLS framework. We have shown that the optimization problem of finding the optimal map is convex and can be solved with standard tools. We have also framed our approach in the literature of DR control, showing the relationship between ambiguity sets. The numerical results show the validity of our findings. Future work encompasses extensions to infinite-horizon control problems, as well as optimal control with state and input constraints.

\bibliographystyle{IEEEtran}
\bibliography{reference}
\appendix
\section{Proofs}
\subsection{Proof of Proposition \ref{prop: relationships}}
\label{appendix: relationships}
We start by proving the first point of the proposition.
\begin{align*}
W_c^{\epsilon}(\Prob, \Q) 
&
\labelrel={myeq:first_equality} 
\begin{multlined}[t]
\inf_{\gamma \in \Gamma(\Prob, \Q)}\ \left\{\int_{\mathcal{Z}\times\mathcal{Z}} c(x, y)\dd\gamma(x, y) \right. \\
\left. + \epsilon\int_{\mathcal{Z}\times\mathcal{Z}}-\log\left(\frac{\dd\Prob(x)\dd\nu(y)}{\dd\gamma(x, y)}\right)\dd\gamma(x, y) \right\}    
\end{multlined}
\\
&
\labelrel\geq{myeq:first_inequality}
\begin{multlined}[t] 
\inf_{\gamma \in \Gamma(\Prob, \Q)}\ \left\{\int_{\mathcal{Z}\times\mathcal{Z}} c(x, y)\dd\gamma(x, y) \right. \\
\left.- \epsilon\log\left(\int_{\mathcal{Z}\times\mathcal{Z}}\dd\Prob(x)\dd\nu(y)\right) \right\}
\end{multlined}
\\
&\labelrel={myeq:second_equality}
\inf_{\gamma \in \Gamma(\Prob, \Q)}\!\!\ \left\{\!\int_{\mathcal{Z}\times\mathcal{Z}} c(x, y)\dd\gamma(x, y)\! \right \} =  W_c(\Prob, \Q)\,,
\end{align*}
where in (\ref{myeq:first_equality}) we have rewritten the logarithm by inverting its argument; the logarithm is concave therefore inequality (\ref{myeq:first_inequality}) comes from a trivial generalization of Jensen's inequality, i.e. given $f, g: \R \rightarrow \R$ with $f$ convex, it holds $\E[f(g(x))] \geq f(\E[g(x)])$; finally (\ref{myeq:second_equality}) follows since $\Prob$ and $\nu$ are probability measures and therefore they integrate to one.
\\
Then, for $\rho \geq 0$, given any $\Q\in\mathcal{P}(\mathcal{Z})$ such that $W_c^{\epsilon}(\Prob, \Q) \leq \rho$ it holds that $W_c(\Prob, \Q) \leq \rho$. This in turn implies $\B_{\rho, \epsilon}(\Prob) \subseteq \B_{\rho}(\Prob) \ \forall \epsilon \geq 0$.
\\
We continue showing the second point. Consider the Sinkhorn discrepancy in (\ref{eq:sinkhorn}) and notice that the KL divergence is nonnegative. Therefore 
\begin{multline*}
\epsilon_1\int_{\mathcal{Z}\times\mathcal{Z}}\log\left(\frac{\dd\gamma(x, y)}{\dd\Prob(x)\dd\nu(y)}\right)\dd\gamma(x, y)\\ \leq \epsilon_2\int_{\mathcal{Z}\times\mathcal{Z}}\log\left(\frac{\dd\gamma(x, y)}{\dd\Prob(x)\dd\nu(y)}\right)\dd\gamma(x, y)\,.
\end{multline*}
Since the infimum preserves monotonicity, we can conclude that $W^{\epsilon_1}_c(\Prob, \Q) \leq W^{\epsilon_2}_c(\Prob, \Q)$ for all $\Q\in \mathcal{P}(\mathcal{Z})$. This in turn implies $\B_{\rho, \epsilon_2}(\Prob) \subseteq \B_{\rho, \epsilon_1}(\Prob)$.\\
Finally, for the third point, when the regularization parameter tends to infinity, $W_c^{\epsilon}(\Prob, \Q)$ is finite if and only if the KL divergence term is identically zero. Therefore, 
\begin{equation*}
H(\gamma|\Prob\times\nu) = H(\gamma|\Prob\times\Q) + H(\Q|\nu)\,,
\end{equation*}
and the last expression is zero if and only if both terms are zero given that they are KL divergences, hence non-negative. Therefore, from the second one we get that $\Q = \nu$ and from the first one that $\gamma = \Prob\times\Q$. We can conclude that $\B_{\rho, \infty}(\Prob)$ is either the singleton $\{\nu\}$ or the empty set. This depends on whether the condition $\int_{\mathcal{Z}\times\mathcal{Z}} c(x, y)\dd\Prob(x) \dd\nu(y) \leq \rho$ is satisfied or not.
\subsection{Proof of Proposition \ref{proposition:convexity}}
\label{app:proof_convexity}
Since (\ref{eq:M>0}), (\ref{eq:LMI2}) $\forall i\in[n]$, and (\ref{eq:LMI1 Schur}) are linear matrix inequalities and the achievability constraint on $\mathbf{\Phi}$ is affine, they are all convex constraints. We therefore focus only on the non-linear constraint (\ref{eq:logdet inequality}).
This constraint is linear in $s_i$ and $\zeta_i$. We proceed to show that the non-linear part $\frac{\lambda\epsilon s}{2}\log\left(\frac{\lambda\epsilon}{2}\right) - \frac{\lambda\epsilon}{2}\log\left|\lambda \left(I + \frac{\epsilon}{2}\Sigma^{-1}\right) - Q\right|$ of \eqref{eq:logdet inequality} is jointly convex in $(\lambda, Q)$ for every $\epsilon\geq 0$. To do so, we define the functions $h:\mathbb{S}^s\rightarrow\R$ and $T:\mathbb{S}^s\rightarrow\mathbb{S}^s$ as follows:
\begin{equation*}
h(Q) = \frac{\epsilon s}{2}\log\left(\frac{\epsilon}{2}\right)- \frac{\epsilon}{2}\log |Q|\,, \quad T(Q) = I + \frac{\epsilon}{2}\Sigma^{-1} - Q\,.
\end{equation*}
Since the log-determinant of a matrix is a concave function \cite{boyd2004convex}, $h$ is convex in $Q$. Similarly, $T$ is affine and therefore convex in $Q$. Hence, the function $g:\mathbb{S}^s\rightarrow\R$ given by
\[g(Q) = h(T(Q)) = \frac{\epsilon}{2}\log\left(\frac{(\frac{\epsilon}{2})^s}{\left|I + \frac{\epsilon}{2}\Sigma^{-1} - Q\right|}\right)\,,\]
is convex, since it represents the composition of an affine and a convex function. We then note that the non-linear part of \eqref{eq:logdet inequality} can be rewritten as
\begin{equation*}
    f(\lambda, Q) = \frac{\lambda\epsilon}{2}\log\left(\frac{\left(\frac{\lambda\epsilon}{2}\right)^s}{\left|\lambda \left(I + \frac{\epsilon}{2}\Sigma^{-1}\right) - Q\right|}\right)\,,
\end{equation*}
and that $f(\lambda, Q) = \lambda g(Q/\lambda)$ is the perspective of the function $g$, see \cite[Section 3.2.6]{boyd2004convex}. Since the perspective of a convex function is also convex, we conclude the proof.

\subsection{Proof of Lemma \ref{thm: duality}}
\label{appendix: duality}
For compactness, we denote by $\xi$ the random variable $\w$ and by $\xi_i$ each realization $\hat{\w}_{[N]}^i$ of it. We first verify that our setup satisfies \cref{assumption}. \crefdefpart{assumption}{assumption1} follows since the set of points in $\R^d$ whose norm is unbounded is a $\nu$ null set. \crefdefpart{assumption}{assumption2} holds since the involved expectation is a Gaussian integral which is convergent for any $\epsilon > 0$. Since any continuous function is also measurable and $\ell$ is quadratic, we have that $\ell$ is measurable as per \crefdefpart{assumption}{assumption3}. Finally, we argue that \crefdefpart{assumption}{assumption4} holds in data-driven scenarios as a regular conditional distribution can be constructed as follows. By the law of total probability, any joint probability distribution $\gamma$ of $(z,\xi)$ can be constructed from the marginal $\hat{\Prob}_n$ of $\xi$ and the conditional distribution $\gamma_\xi$ of $z$ given $\xi=\hat{\xi}_i,\ i\in [n]$, that is, we may write $\gamma = \frac{1}{n}\sum_{i=1}^n \delta_{\hat{\xi}_i}\times\gamma_{\xi}$.
We can now prove the lemma starting with the first point. The feasibility condition in \cite[Theorem 3.1]{sinkhorn} is
\[ \rho +\epsilon \E_{\xi\sim\hat{\Prob}_n}\left[\log\E_{z\sim\nu}\left[e^{-c(\xi, z)/\epsilon}\right]\right]\geq 0\,.\]
We compute the inner expectation as
\begin{align*}
&\begin{multlined}[t]
\E_{z\sim\nu}\left[e^{-c(\xi, z)/\epsilon}\right] = C_d^{-1}\int_{\R^d}\exp\Bigl\{\bigl(-(z-\xi)^{\top}(z-\xi) +
\\
- \frac{\epsilon}{2}(z-m)^{\top}\Sigma^{-1}(z-m)\bigr)/\epsilon\Bigr\}\dd\lambda^d(z)
\end{multlined}
\\
&=
\begin{multlined}[t]
C_d^{-1}\int_{\R^d}\exp\biggl\{\biggl(-\frac{1}{2}z^{\top}\left(\frac{2}{\epsilon}I + \Sigma^{-1}\right)z+
\\
\shoveleft[-0.6cm]{+ 2z^{\top}\left(\frac{\xi}{\epsilon} + \frac{1}{2}\Sigma^{-1}m\right)
- \frac{1}{\epsilon}\|\xi\|^2 - \frac{1}{2}\|m\|^2_{\Sigma^{-1}}\biggr)\biggr\}\dd\lambda^d(z)}
\end{multlined}
\\
&=
\begin{multlined}[t]
\frac{\left(\frac{\epsilon}{2}\right)^{d/2}}{\sqrt{\left|\Sigma + \frac{\epsilon}{2}I\right|}}\exp\left[-\frac{1}{\epsilon}\|\xi\|^2-\frac{1}{2}\|m\|^2_{\Sigma^{-1}}+\right.
\\
\shoveleft[-0.36cm]{
\left.+\frac{1}{\epsilon}\left(\xi + \frac{\epsilon}{2}\Sigma^{-1}m\right)^{\top}\left(I +\frac{\epsilon}{2}\Sigma^{-1}\right)^{-1}\left(\xi + \frac{\epsilon}{2}\Sigma^{-1}m\right)\right]\,,}
\end{multlined}
\end{align*}
where the last equality follows from the standard Gaussian integral result 
\[
\scalebox{.95}{
$\int \! \exp\left(-\frac{1}{2}x^{\top} Ax + c^{\top} x\right)\dd\lambda^d(x) =\\ \sqrt{\frac{(2\pi)^d}{|A|}} \exp\left(\frac{1}{2}c^{\top}A^{-1}c\right)\,.$
}
\]
To conclude the proof of the first point we consider the logarithm of the previous expression scaled by $\epsilon$ and take the expectation with respect to $\hat{\Prob}_n$. In this way, we obtain
\begin{equation*}
\begin{split}
&\frac{\epsilon d}{2}\log\left(\frac{\epsilon}{2}\right) -\frac{\epsilon}{2} \log\left|\Sigma + \frac{\epsilon}{2}I\right| - \frac{\epsilon}{2}\|m\|^2_{\Sigma^{-1}} +
\\
+&\frac{1}{n} \sum_{i=1}^n \star^{\top}\left(I +\frac{\epsilon}{2}\Sigma^{-1}\right)^{-1}\left(\hat{\xi}_i + \frac{\epsilon}{2}\Sigma^{-1}m\right) - \|\hat{\xi}_i\|^2\,.
\end{split}
\end{equation*}
Next, we provide the proof for the second point. From (\ref{eq:dual}) we have the expression for the strong dual for a generic loss function $\ell(\cdot)$. We now focus on the inner expectation. With a quadratic loss function and transport cost, we have 
\begin{equation*}
\begin{aligned}
&\E_{z \sim \nu}\left[e^{(z^{\top} Qz + 2q^{\top}z - \lambda\|\xi-z\|^2)/(\lambda\epsilon)}\right] =
\\
&
\begin{aligned}
\int_{\R^d}\exp\{(z^{\top} Qz &+ 2q^{\top}z + 
\\
&-\lambda(\xi-z)^{\top}(\xi-z))/(\lambda\epsilon)\}\dd\nu(z)=
\end{aligned}
\\
&\begin{aligned}
C_d^{-1}\int_{\R^d}\exp\biggl\{(&z^{\top} Qz + 2q^{\top}z-\lambda(\xi-z)^{\top}(\xi-z))/(\lambda\epsilon)+
\\
&-\frac{1}{2}(z - m)^{\top}\Sigma^{-1}(z - m)\biggr\}\dd\lambda^d(z)=
\end{aligned}
\\
&\begin{aligned}
\alpha\exp\biggl\{\!\biggl[\star^{\top}\! \! &\left(\lambda\left(I +\frac{\epsilon}{2}\Sigma^{-1}\right)-Q\right)^{-1}\! \! \left(q+\lambda\left(\xi+\frac{\epsilon}{2}\Sigma^{-1}m\right)\! \!\right)
\\
& - \lambda \|\xi\|^2 - \frac{\lambda\epsilon}{2}\|m\|^2_{\Sigma^{-1}}\biggr]/(\lambda\epsilon)\biggr\}\,,
\end{aligned}
\end{aligned}
\end{equation*}
with \[\alpha = \frac{\left(\frac{\lambda\epsilon}{2}\right)^{d/2}}{\sqrt{|\Sigma||\lambda\left(I +\frac{\epsilon}{2}\Sigma^{-1}\right)-Q|}}\]
where we require $\lambda\left(I +\frac{\epsilon}{2}\Sigma^{-1}\right) -Q\succ 0$ for the integral to converge.
\\
Then, considering the logarithm of this expression scaled by $\lambda\epsilon$ and taking the expectation with respect to the empirical distribution we obtain
\begin{equation*}
\begin{aligned}
&\inf_{\lambda\left(I+\frac{\epsilon}{2}\Sigma^{-1}\right) \succ Q}\biggl\{\lambda\rho + \frac{1}{n}\sum_{i=1}^n\frac{\lambda\epsilon d}{2}\log\left(\frac{\lambda\epsilon}{2}\right) - \frac{\lambda\epsilon}{2}\log|\Sigma|
\\
&- \frac{\lambda\epsilon}{2}\log\left|\lambda \left(I + \frac{\epsilon}{2}\Sigma^{-1}\right) - Q\right| - \lambda \|\hat{\xi}_i\|^2 - \frac{\lambda\epsilon}{2}\|m\|^2_{\Sigma^{-1}}
\\
&+ \star^{\top}\left(\lambda\left(I +\frac{\epsilon}{2}\Sigma^{-1}\right)-Q\right)^{-1}\left(q+\lambda\left(\hat{\xi}_i+\frac{\epsilon}{2}\Sigma^{-1}m\right)\right)\biggr\}\,.
\end{aligned}
\end{equation*}
To conclude, we can simply introduce epigraphical variables $s_i\in\R$ leading to the final optimization program \eqref{eq:convex program}.

\end{document}